# Obscured AGN Across Cosmic Time

Held at Kloster Seeon, Bavaria, 5 to 8 June 2007

*Bob Fosbury, Carlos De Breuck, Vincenzo Mainieri, Gordon Robertson & Jöel Vernet*

To appear in the ESO Messenger, Sep. 2007

**Affiliations**
Fosbury: ST-ECF
de Breuck, Mainieri, Vernet: ESO
Robertson: University of Sydney & ESO

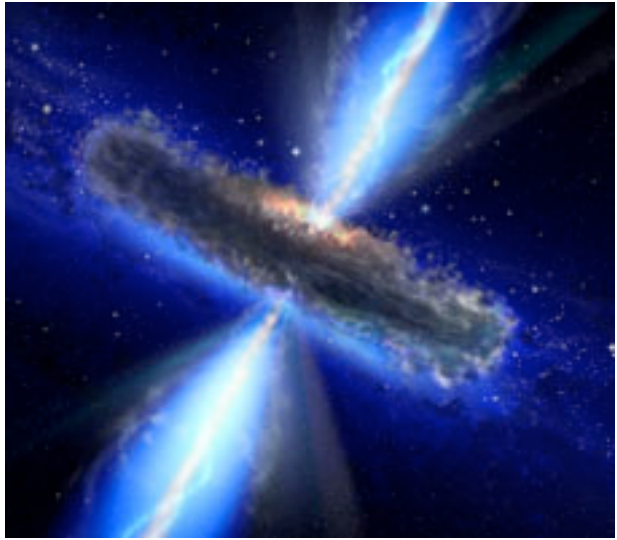

*ESO organises several large international workshops per year. Most of the proceedings have been published in the classical ESO blue series. For this workshop, however, we opted not to publish a printed book. Instead, we made the presentations, including the innovative short 5-minute talks that replaced the posters, available on the conference website (http://www.eso.org/agnii2007). The meeting, attended by 130 people, was held at the Kloster Seeon situated 100 km southeast of Garching in a setting that fostered an especially relaxed and informal atmosphere. This Messenger article provides a necessarily incomplete overview of the many scientific highlights.*

**Abstract** While the radio-loud, obscured quasars (the radio galaxies) have been known and studied for decades, new and sensitive X-ray and mid-infrared surveys are now beginning to reveal large numbers of their radio-quiet counterparts beyond the local Universe. Consequently, we are approaching the compilation of a relatively complete census of AGN of all types coving a large fraction of cosmic time. This is revealing a remarkably intimate connection between the supermassive black hole and its host galaxy. The workshop reported here was designed to explore the results of these rapid observational developments and the nature of the relationships between the stellar and AGN components.

## Introduction

Research areas in astronomy occasionally experience a period of rapid growth due either to the development of some new observational capability or to the simultaneous ripening of several related threads of understanding. In the case of the relationship between supermassive black holes (SMBH) and the growth of their host galaxies, we are in the midst of such a revolution triggered by the combination of both of these effects. While the close connection between the black hole mass and the mass (velocity dispersion) of a galaxy bulge - the 'M-σ' relation - has been known since the late 90's, the mapping out of a comprehensive picture of black hole demographics had to await the deep surveys that were capable of detecting both the obscured and the unobscured flavours of active galaxy: the type 2 and type 1 AGN respectively. The combination of sensitive X-ray and mid-infrared (MIR) observatories in space is now providing just this service by penetrating and/or revealing the obscuration that in type 2 sources hides the nuclear regions in the UV to the near-IR spectrum (see Figure 1 for an example of the X-ray-to-radio SED of a radio galaxy). The results are exciting.

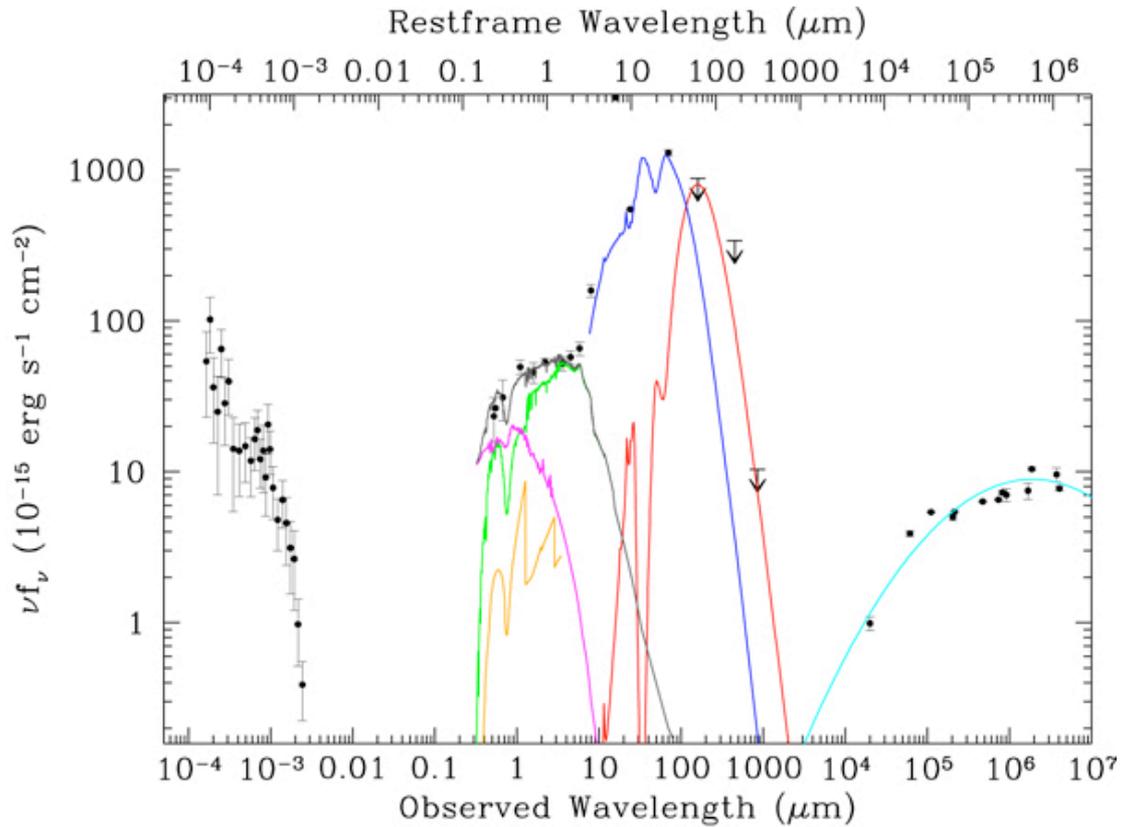

Figure 1: *X-ray to radio spectral energy distribution (SED) of the z=2.483 radio galaxy 4C+23.56 (De Breuck et al. in preparation). All data points are observed flux densities, except for the J, H and K-band fluxes, which have been corrected for strong emission lines. Coloured lines show the decomposition of the SED into different intimately related components. The accretion onto the supermassive black hole creates very energetic X-ray and UV emission and powerful synchrotron radio jets and lobes. In this typical radio loud, type 2 AGN, the direct view to the nucleus is blocked by an obscuring torus absorbing the soft X-ray radiation that is re-emitted as hot thermal dust emission (blue). Radiation escaping through the torus opening is scattered toward the observer providing a unique periscopic view of the nucleus (magenta). It also ionises the ISM producing nebular continuum emission (orange) and, of course, line emission (which is not shown). The massive stellar host galaxy is seen directly in the rest-frame near-IR (green), while an obscured starburst is revealed by sub-mm cool dust re-emission (red). The radio synchrotron fit is in cyan. The bolometric contributions from the accretion onto the supermassive black hole and the stellar nucleosynthesis are remarkably similar.*

The completion of such a census has substantial cosmological significance since it is providing the foundation for identifying the role of AGN feedback in the galaxy formation process. The type 2 sources are of particular value here since, by acting as their own coronagraphs, they facilitate the study of the star formation activity and the investigation of the correlated growth of the black hole and the host galaxy.

While radio galaxies - which are being used to trace the massive galaxy population at all epochs - have been studied intensively for the past 40 years, their more numerous radio quiet counterparts beyond the local universe are only now being discovered in substantial numbers. One of the workshop's aims was to bring together the established radio galaxy community with those studying the radio quiet sources and so help to elucidate the effects of the (possibly) different host galaxies and environment on the manifestation of the AGN phenomenon.

## Advances in capability

Advances in observational capability have been crucial for the rapid development in understanding that we are witnessing. The ability to locate and identify the radio-quiet type 2 objects has depended on the concerted use of both the X-ray and the MIR emissions that derive their energy from the AGN either directly or through re-emission. We have known for a long time that optical and near infrared (NIR) radiations emerge from the AGN in a highly anisotropic manner. However, the X-rays - especially at higher energies - and the MIR are believed to emerge in a more isotropic manner. This enables surveys using these wavebands to find a much more complete set of AGN, regardless of their orientation in space or distribution of their obscuring material.

The sensitivity of the spaceborne instruments is such that active galaxies, at least the more luminous examples, can be seen out to the highest redshifts. The wide field, sensitive groundbased surveys are playing their part as well. The SDSS has revealed very large numbers of AGN (see for example, Figure 4), including the few QSOs with the highest known redshifts. The deep, narrower field, multi-wavelength surveys such as HUDF, GOODS, GEMS, COSMOS etc. are mapping relatively small areas at unprecedented depth and producing samples of fainter AGN. The deepest X-ray surveys are reaching AGN source densities of 7000/deg$^2$, with more than half of these being partially obscured in X-rays. These surveys are allowing, now more than 40 years after its discovery, a huge step forward in the understanding of the origin of the X-ray background (XRB). They confirm the prediction of the population synthesis models which explain the XRB spectrum as the emission, integrated over cosmic time, of obscured and unobscured AGN. Most of the XRB emission below 10 keV is resolved into unobscured and Compton-thin AGN. These sources, however, fall short of matching the XRB peak intensity at 30 keV, which can instead be accounted for by a large (as numerous as that of Compton-thin AGN) population of heavily obscured, Compton-thick objects.

At the very limits of detection, the stacking of sources detected in one waveband in order to characterise the average behaviour in another can be remarkably illuminating. For instance, the stacking of faint MIR sources reveals a hard X-ray spectrum indicative of large absorbing columns.

One obvious question which arises from this new survey capability is whether we discover any qualitatively new type of source? The answer seems to be not really, although there are certainly objects turning up with AGN characteristics in some wavebands that appear rather boring in some of the more traditional windows.

Studies of individual sources benefit from the sheer power of the large groundbased telescopes, more recently coupled to Integral Field Spectrographs (IFS) fed with AO. The astonishing detail with which the hosts are now being studied was one of the clear highlights of the workshop. Kinematic maps, analyses of stellar populations and element abundances from the spatially resolved ISM emission lines are all active new fields. Particularly exciting was the application of the VLTI to imaging the obscuring torus in nearby sources.

## Nature of obscuration

The type 2 (obscured) AGN were always recognised as being a heterogeneous bunch, even after the orientation-dependent unification scheme had become firmly established following the detection, in the 'true' type 2s, of the hidden type 1s in polarized light. Observers in different wavebands contributed to the confusion by using their own set of selection criteria. It was clear that the 'true' type 2s were obscured by some coherent structure that was identified as an equatorial 'torus' that would absorb UV/optical/NIR light over more than half the sky as seen from the central BH. There was always a suspicion however - now amply confirmed by observations - that, lurking within the class, some objects were being obscured instead by larger-scale structures within the host galaxy. Distinguishing these pseudo type 2s is being done using an arsenal of techniques. In contrast to the torus, the extended obscuration usually has a rather small A_V (a few, but enough to hide the BLR). It is also Compton-thin, unlike some (parts) of the tori. A clear indicator of a proper torus, however, is provided by a MIR signature of hot dust over a range of temperatures up to that of dust sublimation. But perhaps the most striking recent result is the spatial resolution in the MIR - using MIDI on the VLTI - of structures that can be identified with the warm torus material in Circinus (Figure 2) and NGC1068. As expected, these are oriented perpendicular to the ionization cones and outflows. In Centaurus A, however, the MIR emission is unresolved and can probably be associated with synchrotron emission from the footprint of the jet.

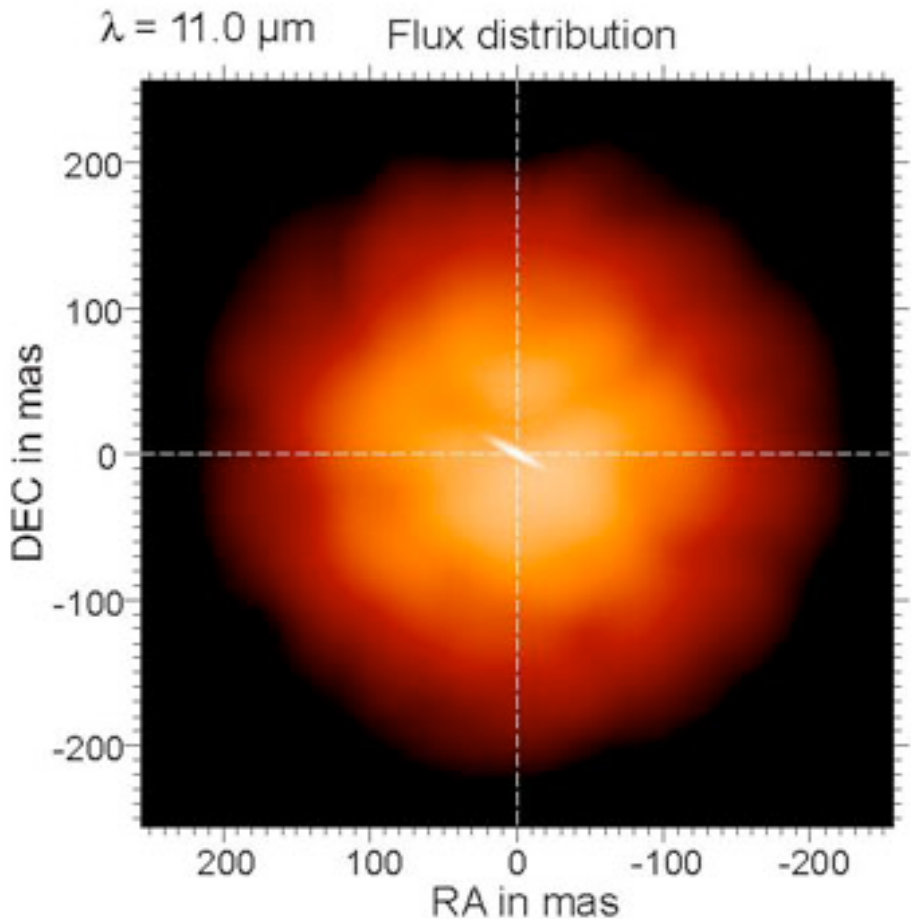

Figure 2: *Flux distribution of the emission from warm dust in the nucleus of the Circinus galaxy. The emission was modelled according to interferometric data obtained with MIDI at the VLTI using two elliptical, Gaussian black body emitters with silicate absorption. The dust is distributed in two components: (1) a small, disk-like component which is oriented perpendicularly to the ionisation cone and outflow and coincides with the orientation and size of a maser disk as well as (2) a larger torus which surrounds the disk component and which shows strong evidence for clumpiness. This finding strongly supports the unified scheme of AGN. (Courtesy Konrad Tristram, tristram@mpia-hd.mpg.de)*

Another fascinating observation, illustrated with X-ray monitoring observations of NGC1365, involves watching a Compton-thin cloud moving at about $10^4$ km/s eclipsing the nuclear X-ray source - having a size of less than $\sim 10^{14}$ cm - over a period of about 2 days. During the eclipse, the broad iron emission line disappears while strong iron absorption lines are seen. This source size corresponds to about 10 Schwarzschild radii for a BH with a mass $\sim 3 \times 10^7$ M_sun.

Our conception of the form of the torus has been derived from many artist's impressions, but what really shapes it? Even a sub-Eddington AGN can produce a radiative force that is comparable to gravity in the material that would comprise a torus. The opacity in the MIR can be some 10 to 30 times that from Thompson scattering, making it possible to support a geometrically thick obscuration. In addition to regular doughnuts, clumpy or otherwise, with radial temperature gradients and gradients within clumps, the old idea of warped disks is still there in the running. We know that such structures exist because we see them, albeit on larger scales, in a number of galaxies.

So what fraction of the type 2s are simply type 1s which happen to be pointing away from us? Well, the prevailing opinion seems to be that about half of them are really of the 'host-obscured' variety that will reveal their AGN to an observer with only a modest degree of determination. There are, however, powerful, deeply obscured sources - ULIRGs - that can fight hard not to reveal their power sources. The debate always used to be: "Are they AGN or Starburst powered?" Now we are more likely to ask about the current balance between these two intimately-related processes.

The astrophysics of all this obscured and obscuring material deep within galaxies has been given a tremendous boost by the availability of exquisite MIR spectroscopy, first with ISO and now with Spitzer (Figure 3). There are many diagnostics in this spectral region, from the permitted and forbidden ionic lines, familiar from their shorter wavelength counterparts, gas-phase molecules like CO, $C_2H_2$, HCN etc., water ice, hydrocarbons (both aromatic and aliphatic) and silicates in dense molecular clouds. Whole new families of 'diagnostic-diagrams' can be constructed in this playground. One example, with silicate absorption/emission strength plotted against the PAH equivalent width, nicely separates different types of AGN and starbursting regions.

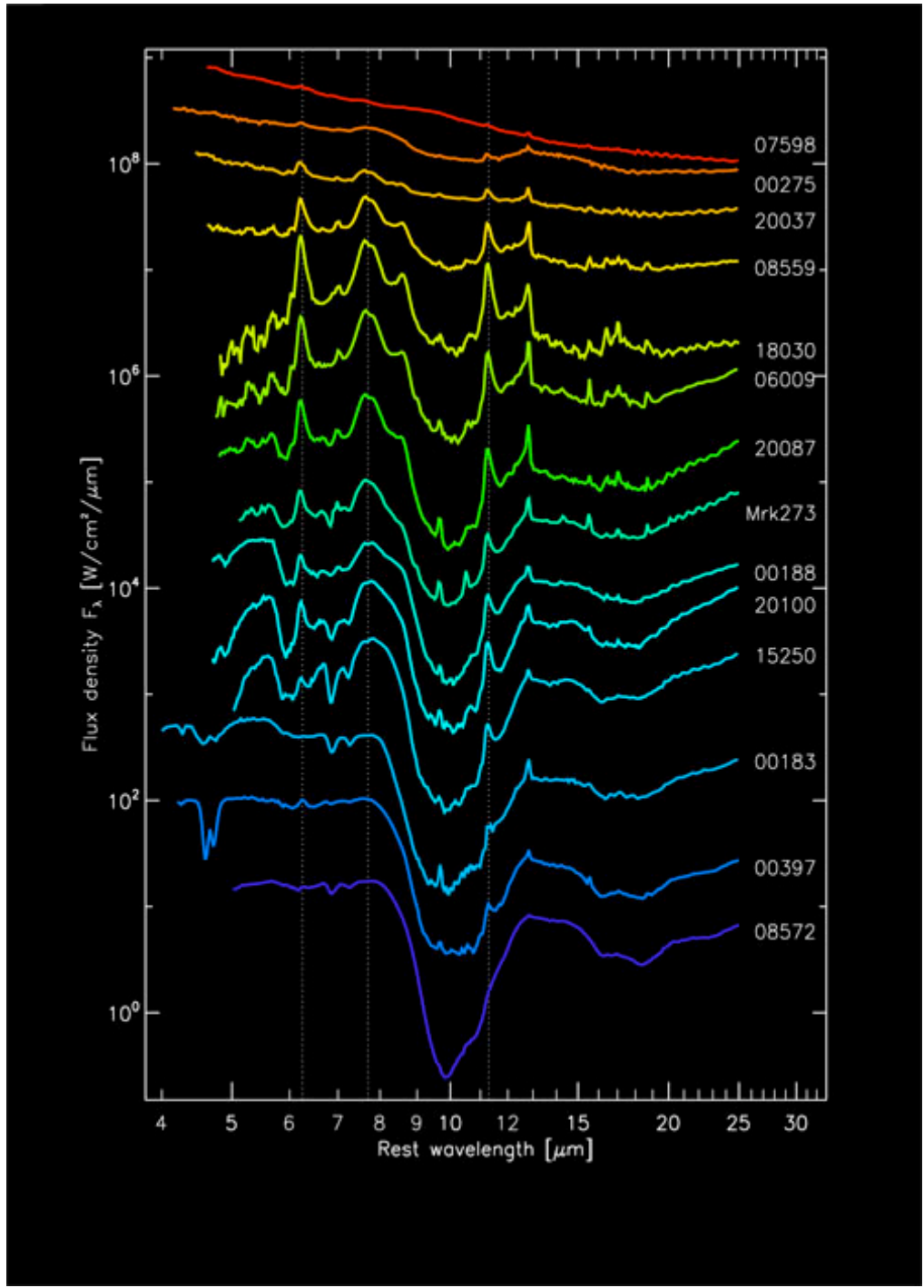

Figure 3: *This collage of low-resolution Spitzer-IRS spectra offers a striking illustration of the diverse nature of the galaxies classified as ultraluminous infrared galaxies (ULIRGs). The reddish spectra at the top are nearly featureless, typical of AGN heated hot dust. Further down, the family of PAH emission features at 6.2, 7.7, 8.6 and 11.3 microns start to appear, indicating an increased contribution of star forming regions to the ULIRG spectrum. The PAH emission starts to dominate in the greenish spectra, while silicate absorption at both 10 and 18 microns becomes apparent. In the spectrum of IRAS 20087, the characteristic absorption edge of water ice at 5.7 microns starts to appear as well, indicating the presence of shielded cold molecular clouds along the line of sight. The water ice feature deepens and the importance of PAH emission decreases moving down to the spectra shown in glacial blue. At the same time, the depth of the 10 and 18 micron silicate features increases and hydrocarbon absorption bands at 6.85 and 7.25 microns become apparent, indicating that the power sources of these ULIRGs are deeply embedded. The bottom three spectra differ from those directly above by the presence of a strong near-infrared continuum and by the relative weakness of the 5.5 to 8.0 micron absorption features. Note how, due to the appreciable redshifts of IRAS 00183 and IRAS 00397, the IRS spectral coverage extends all the way down to rest frame wavelength of 4 microns, facilitating the discovery of wide absorption features due to warm CO gas at 4.6 microns in their spectra. (Spoon et al. 2006; Spoon et al. 2007)*

Examinations of orientation-based unification, eg. the 'Jackson-Browne' test which compares the isotropic emissions from types 1 and 2 sources in matched samples, have been given fresh impetus by the availability of new candidate isotropic AGN emissions in the MIR. By using forbidden line ratios such as [NeV]/[NeII] and [NeV] to low frequency radio power for a matched sample of FRII radio galaxies and quasars, the test is passed with flying colours. This confirms that these type 1 and 2 objects are indeed from the same parent population and also, incidentally, that the [OIII] 5007Å line that was used in the original test is partially obscured in the type 2 objects - as has been suspected for a long time.

## Environments

One of the historical values of AGN has been their utility as 'markers' of distant galaxies and (proto-)clusters. Due to their high visibility, the powerful radio galaxies have long served such a purpose. It is hypothesised, on the basis of their position in the observed K-band Hubble diagram, that these sources indicate the presence of massive hosts. The availability of sensitive MIR measurements from Spitzer has now made it possible to confirm this hypothesis using measurements of the restframe H-band luminosity which is relatively insensitive to AGN contamination and can be easily corrected for the small residuals of it. These sources are also serving as signposts to protoclusters at redshifts beyond the reach of other cluster-finding methods. Multi-band photometry can then enable searches for overdensities of sources showing the characteristics, eg. SED shapes and/or Ly-$\alpha$ emission, expected for 'cluster' members.

Very extended Ly-$\alpha$ halos (>~ 100kpc) indeed appear to be ubiquitous attendants of such objects and they provide information about sources of ionizing radiation and about large-scale gas flows in galaxies early in their evolutionary histories. In addition to those seen around the radio galaxies, Ly-$\alpha$ emission is now being detected around some of the radio quiet sources, especially those seen at sub-mm wavelengths.

## Feedback

The natural theme of the workshop - and indeed the reason for the resurgence of interest in AGN in general - is the rapidly growing evidence for the intimate relationship between the SMBH and the host galaxy, first evinced from the 'M-$\sigma$' relationship. That such a relationship exists was perhaps hinted at by the similarity of the bolometric luminosities originating from nucleosynthesis and from collapse onto black holes through the history of the universe. The actual ratio of these two contributors as a function of cosmic time is something that will be derived as a result of the complete AGN census which we have been describing, but we have known for quite a while that the numbers are roughly comparable. Why should that be so? The physics of energy generation in these two cases is, after all, quite different.

The answer must be that the processes know about one another and are able to communicate by some kind of feedback mechanism. Now, while 'feedback' may be a euphemism for all the physics that we do not understand, we can at least see the results of its operation. AGN seem to be able to switch off cooling flows in clusters and stop star formation in galaxies with the result that the BH mass is tied to that of the host bulge (or vice versa). Indeed, it is remarkable that the cosmic evolution of the star-formation rate density is so closely mirrored by that of the BH accretion rate density. This conclusion has been strengthened recently by several observations. One is the use of a simple, but apparently effective, proxy for BH bolometric luminosity, namely the 'relatively-raw' [OIII] 5007Å luminosity AGN from the Sloan survey (see Figure 4 for an example of a line-ratio diagnostic (BPT) diagram from Sloan data). Another comes from deep X-ray surveys showing that the space density of the less luminous AGN peaks at lower redshifts than that of high X-ray luminosity QSOs. This behaviour is driven by a decrease in the characteristic mass scale of actively accreting black holes as shown by the SDSS data.

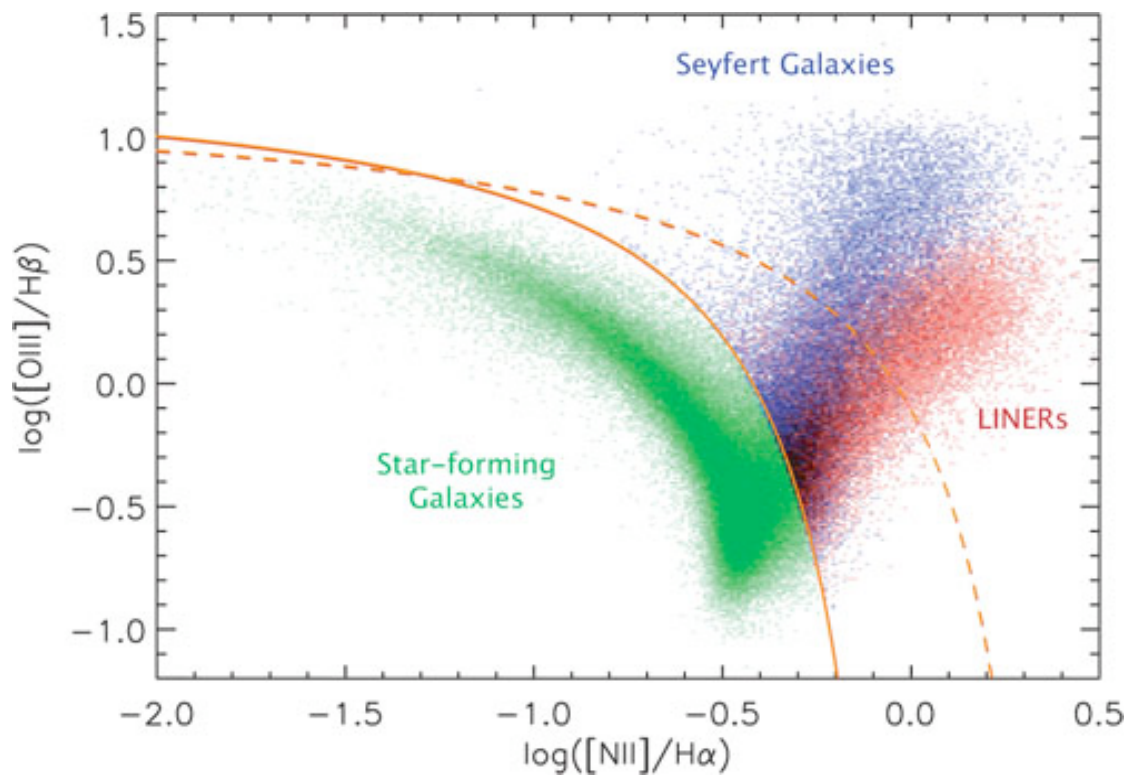

Figure 4: *This figure reveals the spread of emission-line galaxies from the Sloan Digital Sky Survey (SDSS) on the line ratio diagnostic diagram of Baldwin, Phillips & Terlevich (1981). This diagram uses 4 strong optical emission lines, [OIII] 5007Å, [NII] 6583Å, H-α 6563Å, and H-β 4861Å, to distinguish galaxies that are dominated by ionization from young stars (labelled "Star-forming Galaxies"), from those that are ionized by an accreting supermassive black hole in the nucleus (Seyfert and LINER galaxies). The SDSS galaxies not only enable us to see the possible extent of emission line galaxies but also to distinguish the dominant ionization mechanism for this observed spread (Kewley et al. 2006). The curves indicate empirical (solid) and theoretical (dashed) dividing lines between active galactic nuclei (AGN) and star-forming galaxies, based upon the SDSS observations (Kauffmann et al. 2003 and MAPPINGS III photoionization models by Kewley et al. 2002).*

So the 'downsizing' seen in the growth of galaxies is seen also in the accretion rate onto black holes. There are two parts to this problem: one is the question of accretion rate and the associated growth of the BH; the other is the effect of the AGN on the host galaxy and its environment.

But how does the feedback actually work? Understanding this is a demanding problem for both observers and theorists. The feedback can be either negative - by the driving of outflows - or positive - from jet or supernova driven star formation arising as a result of gas compression. There is evidence from several different wavebands for outflows on various scales. The Ly-α halo kinematics also present evidence for inflowing gas. The powerful radio sources appear to be very effective at large (cluster) scales by preventing cooling flows. With their huge mechanical energies, comparable to the bolometric luminosity, these jets can do more than just prevent cooling, they can remove gas entirely: we can see them excavating huge, buoyant cavities within the surrounding halos of hot, X-ray emitting gas. For the lower luminosity AGN, the influence is felt on smaller scales but the BH seem to be able do a good job of controlling the growth of the bulge.

How to grow massive enough BH at early enough times - the Sloan $z \sim 6$ QSO contain SMBH - is not easy to understand. It seems to be possible, just. But there is little room for inefficiency. The wonderfully detailed studies of our own Galactic BH show just how difficult it is to get mass accretion working. Mergers clearly can play a part in getting mass down to the nucleus but this is certainly not the only mechanism. Local AGN appear to be fed routinely from massive disks containing rich reservoirs of cold gas.

## Future

How do we expect the field to develop in the near future? This is always a risky thing to predict. It is safe to say that the surveys will continue to pay a big dividend towards the understanding of the BH demographics over the entire history of the universe. Detailed, sensitive and high resolution observations with the large telescopes will flesh out our primitive understanding of the feedback mechanisms that maintain the balance between stars and BH. We may even reach the point of being able to replace the famous Padovani/Urry AGN cartoon with a real image of the torus and the associated components of the SMBH environment! We also await the capabilities of ALMA to map the cold gas and its motions at small spatial scales to see if this gives insights into the flows of material to the nucleus. Finally, we note that this report is being written just as the observation of the first directly observed (20 year) stellar orbit around our Galactic BH nucleus is being completed.